\documentclass[12pt]{article}
\usepackage{epsf}
\usepackage{cite}
\usepackage{amssymb}
\def\beq{\begin{equation}}

\def\beq{\begin{equation}}
\def\eeq{\end{equation}}
\def\beqn{\begin{eqnarray}}
\def\eeqn{\end{eqnarray}}
\def\bea{\begin{eqnarray}}
\def\eea{\end{eqnarray}} 
\def\be{\begin{equation}}
\def\ee{\end{equation}} 

\begin{document}
\begin{titlepage}
\begin{flushright}MADPH-97-1027
\end{flushright} 
\begin{flushright}December, 1997
\end{flushright}
\vspace{2truecm}
\begin{center}
{\large\bf
QCD CORRECTIONS TO ASSOCIATED HIGGS BOSON PRODUCTION}
\\
\vspace{1in}
{\bf S.~Dawson}\\
{\it Physics Department, Brookhaven National Laboratory,\\
Upton, NY 11973, USA}
\\ 
\vspace{.25in}
{\bf L.~Reina}\\
{\it  Physics Department, University of Wisconsin,\\
Madison, WI 53706, USA}
\vspace{1in}  
\end{center}
\begin{abstract} 
We compute QCD corrections to the processes $e^+e^-\rightarrow t \bar
t h^0$ and $pp\rightarrow t {\overline t}h^0$ by treating the Higgs
boson as a parton which radiates off a heavy quark.  This
approximation is valid in the limits $M_h << M_t$ and $M_h,M_t
<<\sqrt{s}$.  The corrections increase the rate for $pp\rightarrow t
{\overline t}h^0$ at the LHC by a factor of $1.2$ to $1.5$ over the
lowest order rate for an intermediate mass Higgs boson,
$M_h\!<\!180$~GeV.  The QCD corrections are small for $e^+e^-
\rightarrow t \bar t h^0$ at $\sqrt{s}\!=\!1$~TeV.  
\end{abstract}
\end{titlepage}
\clearpage 

\section{Introduction}
The search for the Higgs boson is one of the most important objectives
of present and future colliders.  A Higgs boson or some object like it
is needed in order to give the $W^\pm$ and $Z$ gauge bosons their
observed masses and to cancel the divergences which arise when
radiative corrections to electroweak observables are computed.
However, we have few clues as to the expected mass of the Higgs boson,
which ${\it a~priori}$ is a free parameter of the theory. Direct
experimental searches for the Standard Model Higgs boson at LEP and
LEPII yield the limit,~\cite{leplim}

\beq  
M_h>77.5\,\,\mbox{GeV}\,\,\,\,,
\eeq

\noindent
with no room for a very light Higgs boson.  LEPII will eventually
extend this limit to around $M_h>95$~GeV.  Above the LEPII limit and
below the $2Z$ threshold is termed the \emph{intermediate mass region}
and is the most difficult Higgs mass region to probe experimentally.

In the intermediate mass region, the Higgs boson decays predominantly
to $b {\overline b}$ pairs.  Since there is an overwhelming QCD
background, it appears hopeless to discover the Higgs boson through
the $b {\overline b}$ channel alone.  The associated production of the
Higgs boson could offer a tag through the leptonic decay of the
associated $W$ boson or top quark,\cite{assoc}

\beqn
pp&\rightarrow & Wh^0
\nonumber \\
pp&\rightarrow& t {\overline t} h^0 \,\,\,\,.
\eeqn

\noindent
Both of these production mechanisms produce a relatively small number
of events, making it important to have reliable predictions for the
rates.

The QCD radiative corrections to the process $pp\rightarrow Wh^0$ have
been computed and increase the rate significantly, ($\sim 30-40\% $ at
the LHC) \cite{wb,spirrev}. The QCD radiative corrections to the
processes $e^+e^-\rightarrow t\bar t h^0$ and $pp\rightarrow t
{\overline t} h^0$ do not yet exist and are the subject of this paper.

\begin{figure}[t]
\centering
\epsfxsize=5.in
\leavevmode\epsffile{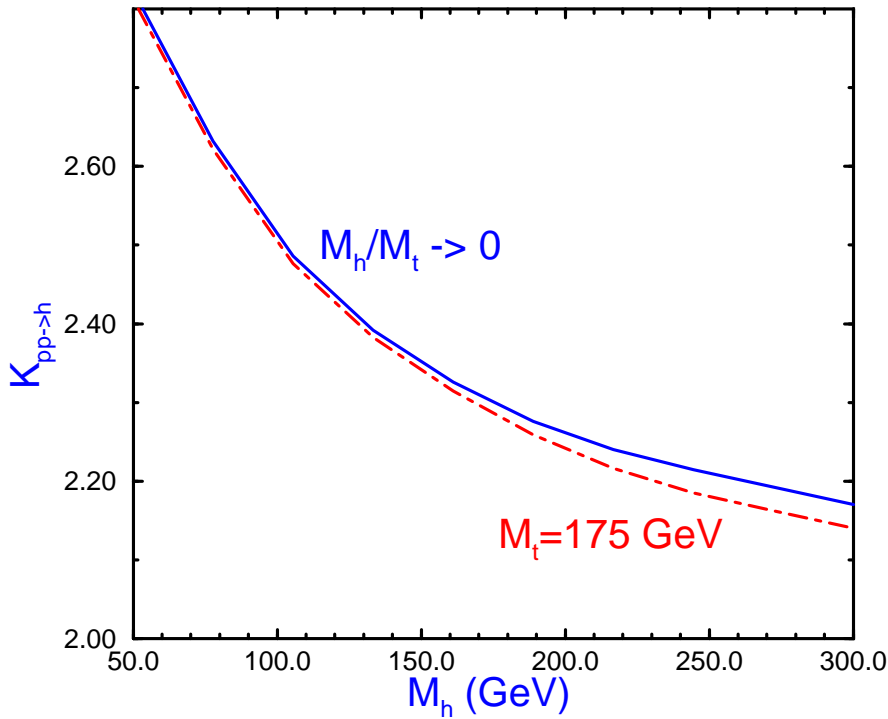}
\caption[]{Ratio of the radiatively corrected cross section
to the lowest order result for $pp\rightarrow h^0$ 
 at the LHC, $\sqrt{s}=14$~TeV, with $\mu=M_h$.
This figure uses the $2-$loop value of $\alpha_s(\mu)$ for both the
lowest order and the NLO predictions.}
\label{kfacfig}
\end{figure}

There is considerable expertise available concerning QCD corrections
to Higgs production. The radiative corrections of ${\cal
O}(\alpha_s^3)$ to $gg\rightarrow h^0$ involve two-loop diagrams and
have been calculated with no approximations \cite{spira}. They have
also been computed in the limit in which $M_h/M_t\rightarrow 0$, where
an effective Lagrangian can be used and the problem reduces to a
one-loop calculation \cite{dawson}. In the intermediate mass region,
one expects this to be a reasonable approximation.  The QCD
corrections to the process $gg\rightarrow h^0$ can be conveniently
described in terms of a $K$ factor,

\beq 
K(\mu)_{pp\rightarrow h}
\equiv {\sigma(pp\rightarrow h^0X)^{NLO}\over 
\sigma(pp\rightarrow h^0)^{LO}}\,\,\,\,, 
\eeq

\noindent where $\mu$ is an arbitrary renormalization scale, which we
take to be $\mu=M_h$.  Fig.~\ref{kfacfig} shows the $K$ factor at the
LHC computed exactly and in the $M_h/M_t\rightarrow 0$ limit.  The
important point is that the $M_h/M_t\rightarrow 0$ limit is extremely
accurate and reproduces the exact result for the $K$ factor to within
$10\%$ for all $M_h < 1$~TeV,\cite{spirrev} although the small
$M_h/M_t$ limit does not accurately give the total cross section for
$M_h>M_t$.  The reason the $K$ factor for this process is so
accurately computed in the small $M_h/M_t$ limit is that a significant
portion of the $K$ factor comes from a constant rescaling of the
lowest order result, along with a universal contribution from the soft
and collinear gluon radiation which is independent of masses \cite{
spira,dawson},

\beq
K(\mu)_{pp\rightarrow h}\sim 1+{\alpha_s(\mu)\over\pi}\biggl(
\pi^2+{11\over 2}+ . . ..\biggr)
\quad .
\eeq

\noindent Hence an accurate prediction for the rate to ${\cal O}(\alpha_s^3)$
can be obtained by calculating the $K$ factor in the large $M_t$ limit
and multiplying this by the lowest order rate computed using the full
mass dependence.

Along this line, we compute the QCD corrections to the processes
$e^+e^-\rightarrow t\bar t h^0$ and $pp\rightarrow t {\overline t}h^0$
in the $M_h/M_t\rightarrow 0$ and high energy limits, which we expect
to be reasonably accurate from our experience with the $gg\rightarrow
h^0$ process.  In these limits the Higgs boson can be treated as a
parton bremsstrahlung off a heavy quark.  Section 2 computes the
distribution of Higgs bosons in a heavy quark, $f_{t\rightarrow h}$,
to leading order in ${M_h\over M_t}$, ${M_t\over \sqrt{s}}$,
${M_h\over \sqrt{s}}$ and to ${\cal O}(\alpha_s)$ in the QCD coupling.
In Section 3, we use the results of Section 2 to compute the QCD
corrections to the process $e^+e^-\rightarrow t {\overline t} h^0$.
We then combine the ${\cal O}({\alpha_s^3})$ corrections to
$pp\rightarrow t {\overline t}$ \cite{nason,smith} with the ${\cal
O}(\alpha_s)$ corrections to $f_{t\rightarrow h}$ to compute the $K$
factor for $pp\rightarrow t {\overline t}h^0$ in the high energy and
large $M_t$ limits.  Just as is the case for the $gg\rightarrow h^0$
process, this $K$ factor can be combined with the lowest order result
computed with the full mass dependences to obtain an accurate estimate
of the radiatively corrected cross section. Section 4 contains some
conclusions.

\section{Higgs  Boson Distribution Function:  The Effective
Higgs Approximation}
\subsection{Lowest Order}  
  
For a Higgs boson which has a mass much lighter than all of the
 relevant energy scales in a given process, $M_h << M_t$, the Higgs
 boson can be treated as a parton which radiates off of the top quark.
 This approach was first proposed by Ellis, Gaillard, and Nanopoulos
 \cite{egn}, and we will refer to it as to the Effective Higgs
 Approximation (EHA). The process $q {\overline q} (gg)\rightarrow t
 {\overline t}h^0$ can be thought of in this approximation as $q
 {\overline q} (gg) \rightarrow t {\overline t}$, followed by the
 radiation of the Higgs boson from the final state heavy quark line.
 This procedure is guaranteed to correctly reproduce the collinear
 divergences associated with the emission of a massless Higgs boson.
 We expect that treating the Higgs boson as a massless parton should
 be a good approximation for the purpose of computing radiative
 corrections and the $K$ factor.

\begin{figure}[t]
\centering
\epsfxsize=3.in
\leavevmode\epsffile{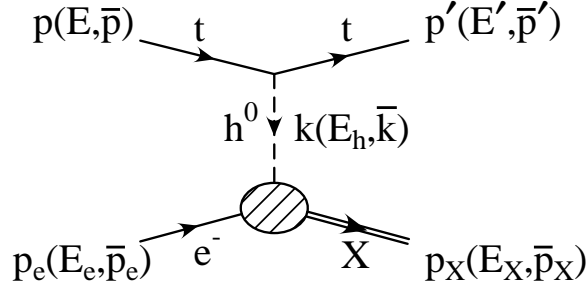}
\caption[]{The scattering process 
$te^-\rightarrow t X$ can be used to extract the Higgs boson structure
function $f_{t\rightarrow h}(x_h)$, assuming that the amplitude is
dominated by the t-channel Higgs pole.}
\label{fhwwfig}
\end{figure}

The distribution function, $f_{t\rightarrow h}$, of a Higgs boson in
a heavy quark, $t$, can be found from the diagram of Fig.~\ref{fhwwfig}
using techniques which were originally derived to find the
distribution of photons in an electron \cite{effp}. This approach has
also been successfully used to compute the distribution of $W$ bosons
in a light quark \cite{effw1}.

The spin- and color-averaged cross section for the scattering process
$t e\rightarrow tX$ through $t-$channel Higgs ($h^0$) exchange is,

\beqn 
\sigma(t e \rightarrow t X)& = & {1\over 48 E E_e} {1\over \mid
{\vec v}_t-{\vec v}_e\mid}\int {{d^3 {\vec p}^{~\prime}} \over (2 \pi)^3}
{1\over 2 E^{\prime}} (2\pi)^4 \delta^{(4)}(k+p_e-P_X) \nonumber \\
&&\qquad \qquad \cdot \mid {\cal A}(t e \rightarrow t X)\mid ^2 d
\Gamma_X\,\,\,\,, 
\eeqn 

\noindent where the kinematics are given in Fig.~\ref{fhwwfig},
${\vec v}_t$ and ${\vec v}_e$ are the velocities of the incoming
particles, and $d\Gamma_X$ is the Lorentz invariant phase space of the
final state $X$.  We factorize the amplitude as

\beq 
\mid {\cal A}(t e \rightarrow t X)\mid^2 = 
\mid {\cal A}(e h^0\rightarrow X)\mid^2 \biggl( {1\over
k^2-M_h^2}\biggr)^2\mid {\cal A}(t\rightarrow t h^0)\mid^2 .
\eeq 

\noindent This factorization is the crucial step in defining the Higgs
distribution function and assumes that the dominant contribution comes
from an on-shell Higgs boson.

We define the energy carried by the Higgs boson to be

\beq
 E_h\equiv x_h E\,\,\,\,,
\label{xh_def}
\eeq

\noindent and neglect all masses, except when they lead to a
logarithmic singularity.\footnote{It is straightforward to include
masses in the analysis.  However, experience with the effective $W$
approximation has shown that the mass corrections are typically
numerically unimportant.} Using

\beq
\sigma(eh^0\rightarrow X)={1\over 8 E_hE_e} 
 d\Gamma_X(2\pi)^4\delta^4(k+p_e-P_X)
\mid {\cal A}(eh^0\rightarrow X)
\mid ^2\quad , 
\eeq

\noindent and $d^3 {\vec p}^{~\prime}=(1-x_h)^2E^3 dx_h d\Omega^\prime$
we find

\beq
\sigma(t e \rightarrow t X)
={E^2\over 96\pi^3} \int dx_h ~x_h (1-x_h)
\int{d\Omega^\prime \over k^4}
\mid {\cal A}(t\rightarrow t h^0)\mid^2 \sigma (e h^0 \rightarrow X)
\quad .
\eeq

We now define the Higgs boson structure function by the relation,

\beq
\sigma(t e \rightarrow t X)\equiv \int f_{t\rightarrow h}(x_h)
\sigma(e h^0\rightarrow X) dx_h 
\quad .  
\eeq 

\noindent This is a Lorentz invariant definition which can be used
 beyond the leading order.  We use the notation $f^0_{t\rightarrow
 h}(x_h)$ to denote the lowest order result,

\beq
f^0_{t\rightarrow h}(x_h)={E^2\over 96\pi^3} x_h (1-x_h) \int
{d\Omega^\prime\over k^4}\mid {\cal A}(t\rightarrow t h^0)
\mid^2    ,
\label{fint}
\eeq
where ${\cal A}(t\rightarrow t h^0)\equiv -i g_t \delta_{ij}
\bar u (p^\prime) u(p)$.

We now consider a Higgs boson which couples to heavy quarks
with the Yukawa interaction,

\beq 
{\cal L}=- g_t {\overline t} t h^0\,\,\,. 
\label{lyuk}
\eeq

\noindent In the Standard Model, $g_t={M_t\over v}$, with $v=246$~GeV.
Our calculation is, however, valid in any model where $M_h << M_t$,
such as a SUSY model.

Performing the integral of Eq.~(\ref{fint}), we obtain our primary
 result,

\beqn
\label{fh_noqcd}
f^0_{t\rightarrow h}(x_h)&=&  {g_t^2\over 16\pi^2} 
\biggl\{ 
{4 (1-x_h)\over x_h}+x_h\log \biggl({ 4 E^2 (1-x_h)^2
\over M_t^2 x_h^2 }\biggr)
\biggr\}
\nonumber \\
&&
+{\cal O}\biggl({M_t\over E}, {M_h\over E}, {M_h\over
M_t}\biggr)\,\,\,\,.
\eeqn

\noindent The dominant numerical contribution comes from the
$(1-x_h)/x_h$ term, i.e. from the \emph{infrared part} of the Higgs
distribution function. In fact, we can also reproduce the first term
in Eq.~(\ref{fh_noqcd}) if we calculate, in the eikonal approximation,
the bremsstrahlung of a \emph{soft} Higgs from the final state of the
$e^+e^-\rightarrow t\bar t$ process. This provides us with an
independent check of the leading behavior of the Higgs structure
function we are going to use in the following.

It is instructive to compute the neglected mass dependent terms using
a different approach.  The cross section for $q {\overline
q}\rightarrow t {\overline t} h^0$ has been computed many years
ago \cite{qqh},

\beqn
{d\sigma(q {\overline q} \rightarrow t
{\overline t} h^0)\over d x_h}
&=& {\alpha_s^2 g_t^2 \over 27 \pi s}
\biggl\{
\biggl[ x_h+2\biggl({4M_t^2-M_h^2\over s}\biggr)
\nonumber \\
&&
+{2\over x_h}
{(4M_t^2-M_h^2)(2M_t^2-M_h^2)\over s^2
}
+{8M_t^2\over s x_h}\biggl]\log\biggl({x_h+
{\hat\beta}\over x_h-{\hat \beta}}\biggr)
\nonumber \\
 && + {4 {\hat \beta}\over x_h^2-{\hat \beta}^2}
\biggl(1+{2M_t^2\over s}\biggr)\biggr({4 M_t^2-M_h^2\over s}
\biggr)\biggr\}\,\,\,\,,
\label{qqresult}
\eeqn  

\noindent where $x_h=2E_h/\sqrt{s}$  is the exact equivalent of the
quantity defined in Eq.~(\ref{xh_def}) and varies in the range between

\be
x_h^{min} = \frac{2M_h}{\sqrt{s}}\,\,\,\,\,\,\,\,\,\mbox{and}
\,\,\,\,\,\,\,\,
x_h^{max} = 1-\frac{4M_t^2}{s}+\frac{M_h^2}{s}\,\,\,\,\,,
\label{xh_intbounds}
\ee

\noindent while $\hat\beta$ is defined to be

\be
\hat\beta=\left[\frac{(x_h^2-(x_h^{min})^2)(x_h^{max}-x_h)}
{(x_h^{max}-x_h+4 M_t^2/s)}\right]^{1/2}\,\,\,\,.
\label{beta}
\ee

We can define the distribution of Higgs bosons in a heavy quark by
the relation,

\beq
\sigma(q {\overline q}\rightarrow t {\overline t} h^0)
=2\int d x_h \sigma(q {\overline q}\rightarrow t {\overline t})
{\hat f}_{t\rightarrow h}(x_h)\,\,\,\,,
\label{fdef} 
\eeq

\noindent where the factor of $2$ reflects the fact that the Higgs
boson can radiate from either quark.  Using Eq.~(\ref{fdef}) we find
an alternate definition of the Higgs distribution function,

\beq 
{\hat f}_{t\rightarrow h}(x_h)
=\biggl[{1\over 2 \sigma( q {\overline q}
\rightarrow t {\overline t})}\biggr]{d\sigma(q {\overline q}
\rightarrow t {\overline t} h)\over dx_h}\,\,\,\,.
\label{fdefnew} 
\eeq

\noindent Taking the limit $s>>M_t^2>>M_h^2$ of
Eq.~(\ref{qqresult}), we find 

\beq
{\hat f}_{t\rightarrow h}(x_h)={g_t^2\over 16 \pi^2}
\biggl\{ x_h\log\biggl({
 s (1-x_h)\over  M_t^2}\biggr)+4\biggl({1-x_h\over x_h}
\biggr)\biggr\}\,\,\,\,.
\label{fh_pptth}
\eeq

\noindent Except for the argument of the logarithm, this agrees with
Eq.~(\ref{fh_noqcd}). It is easy to convince ourselves that this
discrepancy can only introduce a difference of the same order of
magnitude as the effects we are neglecting and is therefore irrelevant
in our approximation. We have explicitely checked that the use of
$f_{t\rightarrow h}(x_h)$ in the form of Eq.~(\ref{fh_noqcd}) or of
Eq.~(\ref{fh_pptth}) is numerically irrelevant for an intermediate
mass Higgs boson and parton sub-energies above around $1$~TeV.

\subsection{QCD Corrections}

The utility of the results of the previous section is that it is
straightforward to compute the ${\cal O}(\alpha_s)$ corrections to
$f_{t\rightarrow h}$.  The discussion here parallels that of
Ref.~\cite{effw2}.

\begin{figure}[t]
\centering
\epsfxsize=6.in
\leavevmode\epsffile{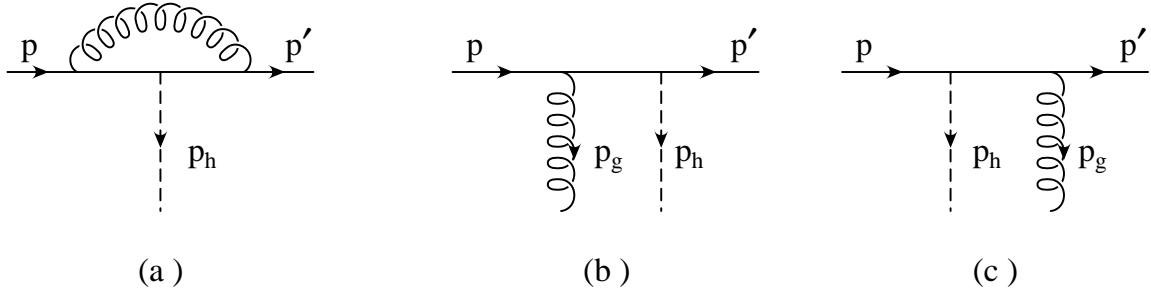}
\caption[]{Virtual (a) and real (b,c) QCD corrections to be computed
in order to obtain the $O(\alpha_s)$ corrections to the structure
function $f_{t\rightarrow h}(x_h)$ in the EHA.}
\label{fhqcdfig}
\end{figure}

\subsubsection{Virtual Corrections} 

The virtual corrections include contributions from the vertex
renormalization of Fig.~\ref{fhqcdfig}a, from the top quark
wavefunction renormalization, and from the top quark mass
renormalization.  The virtual correction from Fig.~\ref{fhqcdfig}a is
given by,

\bea
{\cal A}_V&=&{\cal A}^0(t\rightarrow t h^0)\frac{\alpha_s}{4\pi}C_F\left[
\left(\frac{4\pi\mu^2}{M_t^2}\right)^\epsilon 4(1+2\epsilon)
\Gamma_{UV}(\epsilon) - 6\, \right. \nonumber\\
&&+\left.\left(\frac{4\pi\mu^2}{M_t^2}\right)^\epsilon (2-r_h)
\left(\Gamma_{IR}(\epsilon)+\frac{r_h}{6}(1-\epsilon)\Gamma_{IR}(\epsilon)+
O(r_h^2)\right)\right],
\label{vertex}
\eea

\noindent where we define $r_h=M_h^2/M_t^2$ and $C_F={4\over 3}$.
${\cal A}^0(t\rightarrow t h^0)$ denotes the tree level amplitude for
$t\rightarrow t h^0$, and can be directly read from Eq.~(\ref{lyuk}). We
have explicitely separated the infrared ($\epsilon_{IR}$) and
ultraviolet ($\epsilon_{UV}$) divergences and we have defined

\be
\Gamma(\epsilon) = \frac{1}{\epsilon}-\gamma_E 
\equiv \frac{1}{\epsilon^\prime}
\,\,\,.
\ee

The wave function renormalization ($Z_2$) contributes
to the virtual amplitude,

\beqn 
Z_2 {\cal A}^0(t\rightarrow t h^0) &=& 
-{\cal A}^0(t\rightarrow t h^0)\frac{\alpha_s}{4\pi}
\left(\frac{4\pi\mu^2}{M_t^2}\right)^\epsilon
C_F \biggl[(1+2\epsilon)\Gamma_{UV}(\epsilon)+
2(1+\epsilon)\Gamma_{IR}(\epsilon)\biggr]\nonumber\\
&=& -{\cal A}^0(t\rightarrow t h^0)\frac{\alpha_s}
{4\pi}\left(\frac{4\pi\mu^2}{M_t^2}\right)^\epsilon
C_F \left(\frac{1}{\epsilon_{UV}^\prime}+
\frac{2}{\epsilon_{IR}^\prime}+4\right)
.
\label{wavefunction}
\eeqn 
The sum of (\ref{wavefunction}) and (\ref{vertex}) has both an UV pole
and an IR pole. The UV pole in fact is cancelled by the
renormalization of the mass term in the Higgs-fermion coupling
($g_t=M_t/v$), while the IR pole will be cancelled by the soft gluon
bremsstrahlung contribution (see Fig.~\ref{fhqcdfig}b and
\ref{fhqcdfig}c). To complete the calculation of the virtual part, we
add to (\ref{wavefunction}) and (\ref{vertex}) the mass
renormalization, i.e.

\be
\delta {M_t} = -M_t\frac{\alpha_s}{4\pi}C_F
\left(\frac{4\pi\mu^2}{M_t^2}\right)^\epsilon\Gamma_{UV}(\epsilon)
(3+4\epsilon)\,\,\,\,,
\label{massren}
\ee

\noindent and we obtain,

\be
{\cal A}_{virt}\equiv {\cal A}^0(t\rightarrow t h^0)
\frac{\alpha_s}{4\pi}C_F\left[
-\frac{2}{3}r_h\Gamma_{IR}(\epsilon)
\left(\frac{4\pi\mu^2}{M_t^2}\right)^\epsilon -6 -\frac{1}{3}r_h +
O(r_h^2)\right]
.
\label{amplvirtual}
\ee

Therefore, the virtual processes contribute to the QCD corrected Higgs
distribution function,

\be
f_{t\rightarrow h}^{virt}(x_h) = 
f^0_{t\rightarrow  h}(x_h)\left\{1+\frac{\alpha_s}{2\pi}C_F\left[
-\frac{2}{3}r_h\Gamma_{IR}(\epsilon)
\left(\frac{4\pi\mu^2}{M_t^2}\right)^\epsilon -6 -\frac{1}{3}r_h +
O(r_h^2)\right]\right\}\,\,\,\,.
\label{virtual}
\ee
In the limit $r_h\rightarrow 0$ we have
\be
f_{t\rightarrow h}^{virt}(x_h) \rightarrow 
f^0_{t\rightarrow h}(x_h)\left\{1-\frac{3 \alpha_s}{\pi}C_F
\right\}
.
\label{virtuallim}
\ee
\subsubsection{Real Corrections} 

The gluon bremsstrahlung diagrams are shown in Fig.~\ref{fhqcdfig}b
and \ref{fhqcdfig}c.  In the soft gluon limit, the amplitude is given
by,

\beq
{\cal A}(t\rightarrow t h^0 g)^{soft}=g_s T^a_{ij}
{\cal A}^0(t\rightarrow t h^0)\biggl(
{p^\prime \cdot\epsilon_g\over p^\prime\cdot p_g}
-{p\cdot\epsilon_g \over p\cdot p_g}\biggr)\,\,\,\,.
\eeq

\noindent The contribution to the Higgs distribution function is
found by integrating over the gluon phase space,


\beq
(PS)_g=\int_0^\delta dE_g {E_g^{1-2\epsilon}
\over (2 \pi)^{3-2\epsilon}}
\int_0^\pi \sin^{1-2\epsilon }\theta_1d\theta_1
\int^\pi_0 \sin^{-2\epsilon}\theta_2 d\theta_2,
\label{psglue}
\eeq


\noindent where $E_g$ is the gluon energy.  The integral of 
Eq.~(\ref{psglue}) is cut off at an arbitrary energy scale, $\delta$,
and the dependence on $\delta$ must vanish when the hard gluon terms
are included.  Performing the integrals of Eq.~(\ref{psglue}), the soft
gluon bremsstrahlung term is given by

\be
f_{t\rightarrow h}^{brem}(x_h) =f_{t\rightarrow h}^0(x_h)
\frac{\alpha_s}{2\pi}C_F\left\{
\frac{2}{3}r_h\Gamma(\epsilon_{IR})
\left(\frac{4\pi\mu^2}{\delta^2}\right)^\epsilon +
\Delta+O(r_h^2)\right\}\,\,\,\,,
\label{real}
\ee

\noindent In Eq.~(\ref{real}), $\Delta$ is the sum of all the remaining
finite contributions,

\bea 
\Delta &=&-\frac{2}{3}r_h\log(4)+
\frac{1}{\beta_p}\log\left(\frac{1+\beta_p}{1-\beta_p}\right)+
\frac{1}{\beta_{p^\prime}}\log\left(\frac{1+\beta_{p^\prime}}
{1-\beta_{p^\prime}}\right)\nonumber\\ &&- (2-r_h)\int_0^1
{du\over \beta_v(u)}\frac{\log\left(
\frac{1+\beta_v(u)}{1-\beta_v(u)}\right)}{[1-u(1-u)r_h]} \,\,\,\,,
\label{delta}
\eea 

\noindent and we have denoted by

\be
\beta_p=\left(1-\frac{M_t^2}{E^2}\right)^{1/2}\,\,\,\,\,\,\mbox{and}
\,\,\,\,\,\, \beta_{p^\prime}=\left(1-\frac{M_t^2}{E^{\prime 2}}
\right)^{1/2} 
\ee 

\noindent the 
 velocities of the incoming ($p$) and of
the outgoing ($p^\prime$) t-quark respectively, while

\be 
\beta_v(u) =\left(1-\frac{m_v(u)^2}{E_v(u)^2}\right)^{1/2} 
\ee 

\noindent is the $\beta$-term velocity for the $4-$vector
$v\equiv up+(1-u)p^\prime$, which we use in the integration of the
interference term between the two real emission
diagrams.  In the limit $r_h\rightarrow
0$ and $M_t<<E,E^\prime$,

\be
\Delta\rightarrow \frac{1}{2}\left\{
4 \biggl({E+E^\prime\over E-E^\prime}\biggr)
\log\biggl({E^\prime \over E}\biggr)+8\right\}\,\,\,\,.
\ee

\noindent We can see that in the sum of (\ref{virtual}) and
(\ref{real}) the residual IR divergences cancel.

The final QCD corrected  result is 

\beqn 
\label{fhqcdeha}
f_{t\rightarrow h}(x_h)&=& f_{t\rightarrow h}^0(x_h)
\biggl\{1+{\alpha_s(\mu)\over {\pi}}C_F\biggl[
-1+{2-x_h\over x_h}\log(1-x_h)\biggr]\biggr\} +f^{hard}
+{\cal O}(r_h)
\nonumber \\
&\equiv& f_{t\rightarrow h}^0(x_h)\biggl\{
1+{\alpha_s(\mu)\over \pi}f_{t\rightarrow h}^1(x_h)\biggr\}
\,\,\,\,.
\eeqn

\noindent The hard gluon terms cannot be reproduced in this
approximation and they would require the full calculation of the
bremsstrahlung of a hard gluon in the final state of
$e^+e^-\rightarrow t\bar t h^0$ or $ pp\rightarrow t \bar t
h^0$. Note that to leading order in $r_h$ there is no
dependence on the gluon energy cut-off $\delta$ in the soft
bremsstrahlung contribution of Eq.~(\ref{real}) and so $f^{hard}$ can
contain only finite terms.  In addition,
 there can be no $1/x_h$ singularities
in $f^{hard}$ and so we expect this term to be small. The
emission of a hard gluon should have a very distinguishable
experimental signature and should not affect the study of
$e^+e^-\rightarrow t\bar t h^0$ or $pp\rightarrow t \bar t h^0$.
Therefore we do not include the study of hard gluon emission in this
context and drop the $f^{hard}$ part of the structure function
$f_{t\rightarrow h}(x_h)$.

\begin{figure}[t]
\centering
\epsfxsize=5.in
\leavevmode\epsffile{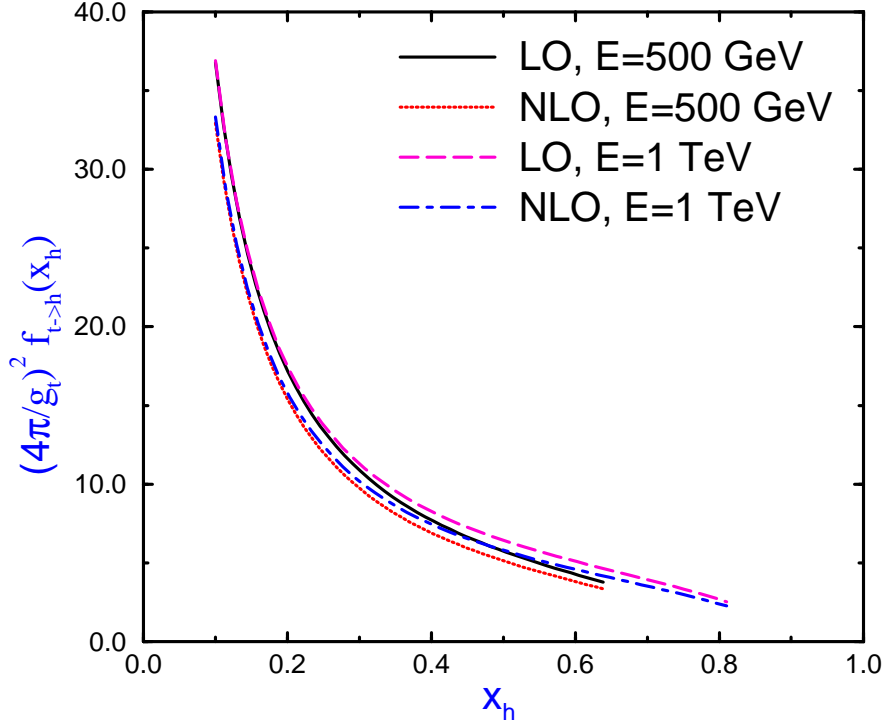}
\caption[]{The Higgs distribution function $f_{t\rightarrow h}(x_h)$ 
in units of $g_t^2/(4\pi)^2$.}
\label{fhfig}
\end{figure}

Figure~\ref{fhfig} shows the lowest order and radiatively corrected
results for 2 typical energy scales.  It is clear that the dominant
contribution is from the $(1-x_h)/x_h$ term with little energy
dependence and that the order ${\cal O}(\alpha_s)$ contributions are
small.

\section{Cross Sections}
\subsection{$e^+e^-\rightarrow t {\overline t} h^0$}

As a first application of the formalism previously derived, we want
now to estimate the impact of QCD corrections on the cross section for
$e^+e^-\rightarrow t\bar t h^0$. This will require us to use the QCD
corrected distribution function derived in Eq.~(\ref{fhqcdeha})
together with the NLO expression for  the cross section for
$e^+e^-\rightarrow t\bar t$.

Let us start testing the validity of the EHA approach in the absence
of QCD corrections. For this purpose, we evaluate and compare the
cross section for $e^+e^-\rightarrow t\bar t h^0$ in the exact theory
(i.e. the Standard Model) and in the EHA, for an intermediate mass
Higgs boson.

The cross section for $e^+e^-\rightarrow t\bar t h^0$ has been first
calculated in \cite{gounaris} (photon-exchange contribution only) and
then completed in \cite{djouadi} (both photon and $Z$-exchange
contributions).  To make the comparison with the formalism explained in
Section 2 easier, we will write the cross section in the form

\be
\sigma(e^+e^-\rightarrow t\bar t h^0) =
\int_{x_h^{min}}^{x_h^{max}}dx_h 
\frac{d\sigma(e^+e^-\rightarrow t\bar t h^0)}{dx_h}\,\,\,\,,
\label{sigma_eetth_exact}
\ee

\noindent where $x_h=2E_h/\sqrt{s}$ and the integration bounds have
been defined in Eq.~(\ref{xh_intbounds}). The differential
distribution in Eq.~(\ref{sigma_eetth_exact}) can be written as

\bea 
\frac{d\sigma(e^+e^-\rightarrow t\bar t h^0)}{dx_h} &=& 
N_c\frac{\sigma_0}{(4\pi)^2}\left\{
\left[Q_e^2Q_t^2+\frac{2Q_eQ_tV_eV_t}{1-M_Z^2/s}+
\frac{(V_e^2+A_e^2)(V_t^2+A_t^2)}{(1-M_Z^2/s)^2}\right]G_1+\right.
\nonumber\\
&&\left.\frac{V_e^2+A_e^2}{(1-M_Z^2/s)^2}\left[A_t^2\sum_{i=2}^6G_i+
V_t^2(G_4+G_6)\right]+\frac{Q_eQ_tV_eV_t}{1-M_Z^2/s}G_6 \right\}
\,\,\,\,,
\label{dsigmadxh_eetth_exact}
\eea

\noindent where $\sigma_0\!=\!4\pi\alpha_e^2/3s$, $\alpha_e$ is the
QED fine structure constant, and  $Q_i$, $V_i$ and $A_i$ ($i\!=\!e$,
$t$) denote the electromagnetic and weak couplings of the electron and
of the top quark respectively,

\be
V_i =\frac{2I_{3L}^i-4Q_is_W^2}{4s_Wc_W}\,\,\,\,\,\,\,\,,\,\,\,\,\,\,\
A_i = \frac{2 I_{3L}^i}{4s_Wc_W}\,\,\,\,,
\ee

\noindent with $I_{3L}^i\!=\!\pm 1/2$ being the weak isospin of the
left-handed fermions and $s_W^2\!=\!1-c_W^2=0.23$.

\noindent The coefficients $G_1$ and $G_2$ describe the radiation of
the Higgs boson off the top quark (both photon and $Z$ boson
exchange), while $G_3,\ldots,G_6$ represent the radiation of the Higgs
boson off the $Z$ boson. Explicit expressions for the coefficients
$G_1,\ldots,G_6$ are given in the Appendix.

As already observed in Ref.~\cite{djouadi}, the most relevant
contributions are those in which the Higgs boson is emitted from a top
quark leg, i.e. those proportional to $G_1$ and $G_2$ in
Eq.~(\ref{dsigmadxh_eetth_exact}). This is further confirmed when we
calculate the same process in the EHA and we see that it is
numerically consistent with the exact result to within a factor of
$1.5-1.8$ for $50\le M_h\le 150$~GeV and $\sqrt{s}\!=\!1$~TeV (the
agreement is clearly improved when we go to higher values of
$\sqrt{s}$).

In fact, according to Section 2, the cross section for
$e^+e^-\rightarrow t\bar t h^0$ is evaluated in the EHA as the
convolution of the Higgs boson distribution function with the cross
section for $e^+e^-\rightarrow t\bar t$, i.e.

\be
\sigma(e^+e^-\rightarrow t\bar t h^0)_{EHA} = 
2 \int_{x_h^{min}}^{x_h^{max}}dx_h f^0_{t\rightarrow h}(x_h)
\sigma(e^+e^-\rightarrow t\bar t)\,\,\,\,,
\label{sigma_eetth_eha1}
\ee

\noindent and therefore only the emission of a Higgs boson from a top
quark is taken into account. 

\noindent In the absence of QCD corrections, $f^0_{t\rightarrow
h}(x_h)$ is given by Eq.~(\ref{fh_noqcd}), while
$\sigma(e^+e^-\rightarrow t\bar t)$ can be easily calculated and reads
\cite{jersak,harlander}

\be
\sigma(e^+e^-\rightarrow t\bar t) = 
\beta_t\left(1+\frac{1}{2}\mu_t^2\right)\sigma_{VV}+\beta_t^3\sigma_{AA}
\,\,\,\,,
\label{sigma_eett}
\ee

\noindent where $\mu_t\!=\! 2M_t/\sqrt{s}$,
$\beta_t^2\!=\!(1-\mu_t^2)$, and $\sigma_{VV}$ and $\sigma_{AA}$
correspond respectively to the product of two vector or two axial
currents (the interference between the two gives zero upon angular
integration),

\bea
\label{sigma_eett_vv_aa}
\sigma_{VV} &=& N_c \sigma_0\left[
Q_e^2Q_t^2 + 2Q_eQ_tV_eV_t \frac{s}{s-M_Z^2}+
V_t^2(V_e^2+A_e^2)\left(\frac{s}{s-M_Z^2}\right)^2 \right]\nonumber \\
\sigma_{AA} &=& N_c\sigma_0\,A_t^2(V_e^2+A_e^2)
\left(\frac{s}{s-M_Z^2}\right)^2\,\,\,\,,
\eea

\noindent using the same notation we introduced before. We notice that
$\sigma(e^+e^-\rightarrow t\bar t)$ does not depend on $x_h$ and the
associated Higgs production, in the EHA, is obtained by simply
multiplying $\sigma(e^+e^-\rightarrow t\bar t)$ by a prefactor, as can
be seen from Eq.~(\ref{sigma_eetth_eha1}).

\begin{figure}[t]
\centering
\epsfxsize=5.in
\leavevmode\epsffile{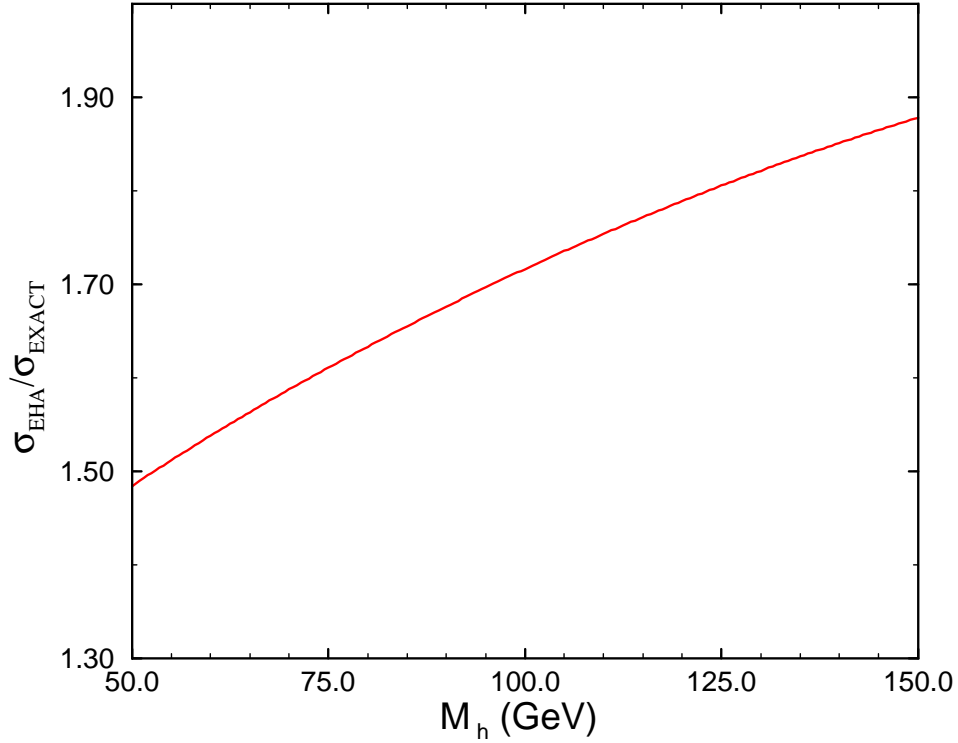}
\caption[]{Ratio of the  lowest order cross section computed
in the EHA for $e^+e^-\rightarrow t\bar t h^0$ to the same quantity
computed exactly, for intermediate mass Higgs bosons at
$\sqrt{s}=1$~TeV.}
\label{ratio_exact_eha}
\end{figure}

\noindent In Figure~\ref{ratio_exact_eha} we compare the exact cross
section of Eq.~(\ref{sigma_eetth_exact}) to the EHA one of
Eq.~(\ref{sigma_eetth_eha1}), in terms of their ratio

\be
R = \frac{\sigma(e^+e^-\rightarrow t\bar t h^0)_{EHA}}
{\sigma(e^+e^-\rightarrow t\bar t h^0)_{EXACT}}\quad .
\label{ratio}
\ee

\noindent As we can see, the EHA reproduces the exact cross section
to a good level of approximation and we are entitled to use it in the
estimate of the impact of QCD corrections on the associated production
of an intermediate mass Higgs boson in $t\bar t$ events.

 Within the EHA formalism, we can parameterize the effect of
QCD corrections as 

\be
\sigma(e^+e^-\rightarrow t\bar t h^0)=
2\left[\int_{x_h^{min}}^{x_h^{max}} dx_h 
f^0_{t\rightarrow h}(x_h)\left(1+\frac{\alpha_s(\mu)}{\pi}
f^1_{t\rightarrow h}(x_h)\right)\right] 
\left(\hat\sigma^0+\frac{\alpha_s(\mu)}{\pi}\hat\sigma^1\right)
\,\,\,\,,
\label{sigma_eetth_eha_QCD}
\ee

\noindent where $f^0_{t\rightarrow h}$ is the Higgs boson distribution
function without QCD corrections, as in Eq.~(\ref{fh_noqcd}), while
the exact expression of $f^1_{t\rightarrow h}(x_h)$ can be derived
from Eq.~(\ref{fhqcdeha}).  Moreover, we have schematically
represented the QCD corrected cross section $\sigma(e^+e^-\rightarrow
t\bar t)$ in the form

\be
\sigma(e^+e^-\rightarrow t\bar t) = \hat\sigma_0+
\frac{\alpha_s(\mu)}{\pi}\hat\sigma^1\,\,\,\,,
\label{sigma_eett_QCD}
\ee
where $\hat\sigma_0$ is given in Eq.~(\ref{sigma_eett}).

\noindent Numerical expressions for $\hat\sigma^1$
at various energies can be found in the literature \cite{harlander}
and we refer to these numbers in our discussion. The impact of QCD
corrections on the prediction of the cross section for
$e^+e^-\rightarrow t\bar t h^0$ is described by the K-factor,

\be
K(\mu)_{e^+e^-\rightarrow t\bar t h}
 = \frac{\sigma(e^+e^-\rightarrow t\bar t h^0)^{NLO}}
{\sigma(e^+e^-\rightarrow t\bar t h^0)^{LO}} \quad .
\ee

\noindent In the EHA, using Eq.~(\ref{sigma_eetth_eha1}) and
Eq.~(\ref{sigma_eetth_eha_QCD}), the K-factor can be written as

\bea
K(\mu)_{e^+e^-\rightarrow t\bar t h} &=& \frac{
\left[\int_{x_h^{min}}^{x_h^{max}} dx_h 
f^0_{t\rightarrow h}(x_h)\left(1+\frac{\alpha_s(\mu)}{\pi}
f^1_{t\rightarrow h}(x_h)\right)\right] 
\left(\hat\sigma^0+\frac{\alpha_s(\mu)}{\pi}\hat\sigma^1\right)}
{\left[\int_{x_h^{min}}^{x_h^{max}} dx_h 
f^0_{t\rightarrow h}(x_h)\right]\hat\sigma^0}\nonumber\\
&=& 1+\frac{\alpha_s(\mu)}{\pi}\left(\frac{\hat\sigma^1}{\hat\sigma^0}+
\frac{\int_{x_h^{min}}^{x_h^{max}} dx_h 
f^0_{t\rightarrow h}(x_h)f^1_{t\rightarrow h}(x_h)}
{\int_{x_h^{min}}^{x_h^{max}} dx_h 
f^0_{t\rightarrow h}(x_h)}\right)\,\,\,\,.
\eea

\noindent Using the results of Ref.~\cite{harlander}, we can estimate
that $\hat\sigma^1/\hat\sigma^0\!\simeq\!1.7$ at $\sqrt{s}\!=\!1$~TeV
(taking $\mu=\sqrt{s}$ and $\alpha_s(s)\!=\!0.088$),
 while the second term in parenthesis is  roughly
$-4$ for $M_h< 180$ GeV.
 Therefore $K(s)\!\simeq
.94$ and the impact of QCD corrections is estimated in the EHA to be
very mild for $e^+e^-$ initial states.

\subsection{$p p \rightarrow t {\overline t} h^0$}

The QCD corrections to the process $pp\rightarrow t {\overline t} h^0$
can be estimated using the EHA and the results of
Ref.~\cite{nason,smith} for the NLO corrections to $pp\rightarrow t
{\overline t}$.  The NLO cross section for $t {\overline t}$
production at a hadron collider can be conveniently parameterized at
the parton level as,

\beq
\sigma(ij\rightarrow t {\overline t}X)=
{\alpha_s^2(\mu)\over M_t^2}\biggl\{ f_{ij}^0({\hat \rho}
)+4 \pi \alpha_s(\mu)
\biggl[ f_{ij}^1({\hat\rho})
+{\overline f}_{ij}^1({\hat \rho})
\log\biggl({\mu^2\over M_t^2}\biggr)\biggr]\biggr\}\,\,\,\,, 
\eeq

\noindent where $ij$ corresponds to the $gg$, $q {\overline q}$, $qg$,
and ${\overline q} g$ initial states and

\beq
{\hat \rho}={4 M_t^2\over {\hat s}}\,\,\,\,,
\eeq

\noindent with ${\hat s}$ the parton sub-energy.  Analytic results for
the $f_{ij}^0({\hat \rho} )$ and 
${\overline f}_{ij}^1({\hat \rho})$ and numerical parameterizations of
$f_{ij}^1({\hat \rho})$  can be
found in Ref.~\cite{nason}.

In the EHA, the cross section for $pp\rightarrow t {\overline t} h^0$
production at lowest order is then,

\beq
\sigma^0(pp\rightarrow t \bar t h^0)
={2 \alpha_s^2(\mu)\over M_t^2}\int dx_1 dx_2 
\sum_{ij} F_i(x_1) F_j(x_2)  f_{ij}^0({\hat \rho})\int dx_h
f_{t\rightarrow h}^0(x_h) ,
\eeq

\noindent where $F_i(x)$ are the parton distribution functions and
 $x_1$ and $x_2$ are the momentum fractions carried by the incoming
 partons. Again, the over-all factor of $2$ reflects the fact that the
 Higgs boson can be radiated off either top quark leg.  The parton
 sub-energy  is ${\hat s} =x_1 x_2 s$.  Note that because
 of the dependence on ${\hat s}$ in $f_{t\rightarrow h}(x_h)$, the
 effect of the Higgs emission is not a simple prefactor as was the
 case for $e^+e^-$, (although the energy dependence is very small).

At lowest order, the process $pp\rightarrow t {\overline t}h^0$
 includes both the $q {\overline q}$ and $gg$ initial states.  As was
 the case for $e^+e^-$, the contribution from the $q {\overline q}$
 initial state is well approximated at lowest order in the
 intermediate mass Higgs region by the EHA calculation.  However, the
 process $gg\rightarrow t {\overline t} h^0$ also contains the
 $t$~-channel emission of a Higgs boson which is not included in the
 EHA approximation and so the EHA is a much poorer approximation to
 the lowest order hadronic cross section than it is to the $e^+e^-$
 cross section.\footnote{The agreement between the exact calculation
 and the EHA calculation of $pp\rightarrow t \bar t h^0$ at the LHC
 can be improved by using Eq.~(\ref{fh_pptth}) instead of
 Eq.~(\ref{fh_noqcd}) for $f_{t\rightarrow h}^0$.  Even so, the EHA
 overestimates the exact cross section for $M_h\sim 150$~GeV by about
 a factor of $2$.  The $K$ factors obtained using Eq.~(\ref{fh_noqcd})
 or Eq~(\ref{fh_pptth}), along with Eq.~(\ref{fhqcdeha}), are,
 however, almost identical.}  We will therefore use the EHA only to
 calculate the $K$ factor. The best estimate of the rate for
 $pp\rightarrow t \bar t h^0$ will then be obtained by using the exact
 calculation for the lowest order rate and then multiplying by the $K$
 factor obtained using the EHA.

The cross section to NLO can easily be found using the EHA,

\beqn
\sigma^1(pp\rightarrow t {\overline t}h^0)
&=&
{2\alpha_s^2(\mu)\over M_t^2}
\int dx_1 dx_2 \sum_{ij} F_i(x_1) F_j(x_2)\int
 dx_h
f^0_{t\rightarrow h}(x_h)\biggl\{f_{ij}^0(\hat\rho)
\nonumber \\
&&+{ \alpha_s(\mu)
\over \pi}
\biggl( f^1_{t\rightarrow h}(x_h)
 + 
(4 \pi^2) \biggl [ f_{ij}^1(\hat\rho)
+{\overline f}_{ij}^1(\hat\rho)
\log\biggl({\mu^2\over M_t^2}\biggr)\biggl]
\biggr)\biggr\}\,\,\,\,.  
\eeqn

The $K$ factor is then given as usual by,

\beq
K(\mu)_{pp\rightarrow t {\overline t}h}
\equiv {\sigma^1(pp\rightarrow t {\overline t}h^0)
\over \sigma^0(pp\rightarrow t {\overline t}h^0)}\,\,\,\,.
\eeq 

\begin{figure}[tb] 
\centering
\epsfxsize=5.in
\leavevmode\epsffile{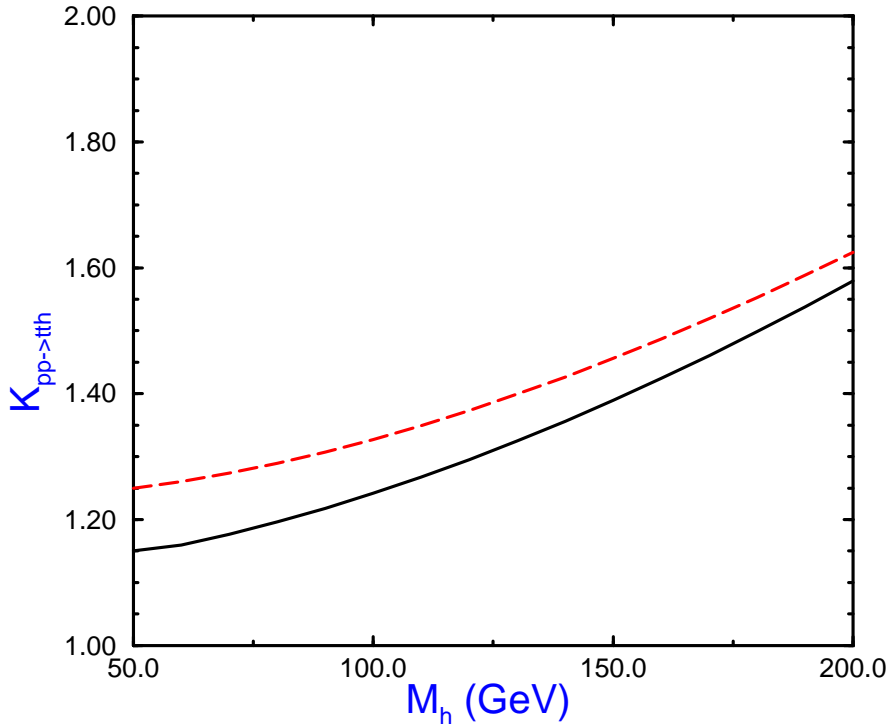}
\caption[]{K factor for $pp\rightarrow t\bar t h^0$ production at
the LHC, $\sqrt{s}=14$~TeV, for $\mu=M_t$, using Morfin-Tung [18]
 structure functions.  The solid curve uses the $1-$ loop evolution of
 $\alpha_s(\mu)$ and lowest order structure functions for $\sigma^0$,
 and the $2-$loop evolution of $\alpha_s(\mu)$ and NLO structure
 functions for $\sigma^1$. The dashed curve uses the $2-$loop
 evolution of $\alpha_s(\mu)$ and the LO structure functions for both
 $\sigma^0$ and $\sigma^1$.}
\label{kfac_pp}
\end{figure}

\noindent In Fig.~\ref{kfac_pp}, we show the $K$ factor 
using structure functions derived by Morfin and Tung.\cite{mt} It is
clear from this figure that using the lowest order cross section with
the $2-$loop evolution of $\alpha_s(\mu)$ overestimates the size of
the corrections.  (This is also true for the $pp\rightarrow h^0$
process.\cite{spirrev}) The $K$ factor varies between around $1.2$ and
$1.5$ for the intermediate mass Higgs boson.  Since the dominant
production mechanism is gluon fusion, the exact value of the $K$
factor is sensitive to the choice of structure functions.  We have
found that changing the set of structure functions induces a
$10-20~\%$ uncertainty in the $K$ factor shown in Fig.~\ref{kfac_pp}.

\section{Conclusions}

We have computed the next-to-leading order QCD corrections to $e^+e^-
\rightarrow t \bar t h^0$ and to $pp\rightarrow t \bar t h^0$ in the
high energy limit and the limit in which $M_h<<M_t$. We expect
this to be a reasonable approximation in the intermediate mass
region. The corrections to $e^+e^-\rightarrow t \bar t h^0$ are small
and can safely be neglected.  The corrections to $pp\rightarrow t \bar
t h^0$, however, increase the rate at the LHC by a factor of between
$1.2$ to $1.5$ for $M_h < 180$~GeV, although the numbers are sensitive
to the choice of structure functions.  Within the context of the EHA,
the bulk of these corrections can be identified with the ${\cal
O}(\alpha_s^3)$ corrections to the $pp\rightarrow t \bar t$
sub-process.  The significant size of the corrections underscores the
need for a complete calculation.

\section*{Acknowledgments}
We are grateful to F.~Paige and D.~Zeppenfeld for valuable discussions
and to M. Spira for a careful reading of the manuscript.  We thank
A.~Stange for providing us with his FORTRAN code with the lowest order
cross section for $p p \rightarrow t \bar t h^0$.  The work of
S.~D. is supported by the U.S.  Department of Energy under contract
DE-AC02-76CH00016. The work of L.~R. is supported by the U.S.
Department of Energy under contract DE-FG02-95ER40896.

\newpage
\appendix
\section{Coefficients for $\sigma(e^+e^-\rightarrow t\bar t h^0)$}

As we have seen in Eq.~(\ref{dsigmadxh_eetth_exact}), the differential
cross section for $e^+e^-\rightarrow t\bar t h^0$ can be expressed in
terms of six coefficients $G_1,\ldots,G_6$. Two of them, $G_1$ and
$G_2$ account for the emission of a Higgs from a top quark and are
explicitely given by

\bea
\label{th_coeff}
G_1&=&\frac{2\,g_t^2}{s^2\left(\hat\beta^2 - x_h^2 \right)x_h}
     \left(\phantom{\frac{1}{2}}\!\!\!\!
        -4\hat\beta\,\left( 4 M_t^2 - M_h^2 \right) \,
        \left( 2 M_t^2 + s \right) x_h + \right.\\
&& \!\!\!\!\!\!\!\!\!\!\!\!\!\!\!\!\!
        \left.\left(\hat\beta^2 - x_h^2 \right) 
        \left( 16M_t^4 + 2M_h^4 - 
          2M_h^2 s x_h + s^2 x_h^2 - 4 M_t^2  
        \left( 3 M_h^2 - 2 s - 2 s x_h \right)  \right) \,
        \log \left({\frac{x_h+\hat\beta}{x_h-\hat\beta}}\right)
       \right)\,\,\,, \nonumber\\
G_2&=& \frac{-2\,g_t^2}{s^2\,\left(\hat\beta^2 - x_h^2 \right) x_h}
     \left(\phantom{\frac{1}{2}}\!\!\!\! \hat\beta\, x_h
        \left( -96 M_t^4 + 24 M_t^2 M_h^2 - 
          s\left( -M_h^2 + s + s x_h \right) 
           \left( -\hat\beta^2 + x_h^2 \right)  \right)\right.  +
     \nonumber\\
&& \!\!\!\!\!\!\!\!\!\!\!\!\!\!\!\!\!
        \left.   2\left( \hat\beta^2 - x_h^2 \right) 
        \left( 24 M_t^4 + 2\left( M_h^4 - M_h^2 s x_h \right)  + 
          M_t^2\left( -14 M_h^2 + 12 s x_h + s x_h^2 \right)  \right) 
        \log\left(\frac{x_h+\hat\beta}{x_h-\hat\beta}\right)
       \right) \,\,\,,\nonumber
\eea

\noindent while the other four coefficients, $G_3,\ldots,G_6$ describe
the emission of a Higgs from the $Z$-boson and can be written in the
following form,

\bea
\label{Zh_coeff}
G_3&=& \frac{-2\,\hat\beta g_Z^2 M_t^2}
{M_Z^2 \left( M_h^2 - M_Z^2 + s - s x_h \right)^2}
     \left( 4 M_h^4 + 12 M_Z^4 + 2 M_Z^2 s x_h^2 + 
       s^2\left( -1 + x_h \right) x_h^2 - \right.\nonumber\\
   && \left.   M_h^2\left( 8 M_Z^2 + s\left( -4 + 4 x_h + x_h^2 \right) 
           \right)  \right)\,\,\,, \nonumber\\
G_4 &=& \frac{\hat\beta g_Z^2 M_Z^2}
 {6\left( M_h^2 - M_Z^2 + s - s x_h \right)^2}
     \left( 48 M_t^2 + 12 M_h^2 - s\left( -24 + \hat\beta^2 + 24 x_h - 
          3 x_h^2 \right)  \right)\,\,\,, \\
G_5 &=&
\frac{-4\,g_t\,g_Z\,M_t}
{M_Z\,s\left( -M_h^2 + M_Z^2 + s\left( -1 + x_h \right)\right) }
     \left(\phantom{\frac{1}{2}}\!\!\!\! \hat\beta\,s  \left( 6 M_Z^2 + 
     x_h\left( -M_h^2 - s + s x_h \right)\right) + \right.\nonumber\\
  && \left.   2\left( M_h^2\left( M_h^2 - 3M_Z^2 + s - s x_h \right)  + 
          M_t^2\left( -4 M_h^2 + 12 M_Z^2 + s x_h^2 \right)  \right) \,
        \log\left({\frac{x_h+\hat\beta}{x_h-\hat\beta}}\right)
       \right)\,\,\,, \nonumber\\
G_6 &=&
\frac{8\,g_t\,g_Z\,M_t\,M_Z}
{s\left( -M_h^2 + M_Z^2 + s\,\left( -1 + x_h \right)\right)}
     \left( \hat\beta\,s + \left( 4 M_t^2 - M_h^2 + 2 s - s x_h \right) 
        \log\left({\frac{x_h+\hat\beta}{x_h-\hat\beta}}\right)
       \right) \nonumber\,\,\,\,.
\eea

\noindent In both Eq.~(\ref{th_coeff}) and Eq.~(\ref{Zh_coeff}), 
$g_t$ and $g_Z$ denote the Yukawa couplings of the top quark
($g_t\!=\!M_t/v$) and of the $Z$ boson ($g_Z\!=\!M_Z/v$) respectively,
while the quantity $\hat\beta$ is defined in Eq.~(\ref{beta}).
Ref. \cite{djouadi} presents the result for $d\sigma/dx_1dx_2$, where
$x_1$ and $x_2$ are the fractional energies of the top quarks.  After
making a change of variables and integrating over $\mid x_1-x_2\mid$,
their expressions for the $G_i$ agree with ours.
\end{document}